\documentclass[1p, numbers, final,times]{elsarticle}
 \usepackage{graphicx}
\usepackage{amssymb}
\journal{Chemical Physics Letters}

\begin{document}

\begin{frontmatter}

%% Title, authors and addresses

%% use the tnoteref command within \title for footnotes;
%% use the tnotetext command for theassociated footnote;/home/gavardi/Documents/
%% use the fnref command within \author or \address for footnotes;
%% use the fntext command for theassociated footnote;
%% use the corref command within \author for corresponding author footnotes;
%% use the cortext command for theassociated footnote;
%% use the ead command for the email address,
%% and the form \ead[url] for the home page:
%% \title{Title\tnoteref{label1}}
%% \tnotetext[label1]{}
%% \author{Name\corref{cor1}\fnref{label2}}
%% \ead{email address}
%% \ead[url]{home page}
%% \fntext[label2]{}
%% \cortext[cor1]{}
%% \address{Address\fnref{label3}}
%% \fntext[label3]{}

\title{A kinetic Monte Carlo study of desorption of H$_2$ from
 graphite (0001)}

%% use optional labels to link authors explicitly to addresses:
%% \author[label1,label2]{}
%% \address[label1]{}
%% \address[label2]{}

\author[milan,leiden]{E.~Gavardi}
\author[leiden]{H.~M.~Cuppen\corref{cor1}}
\author[aarhus]{L.~Hornek\ae r}

\address[milan]{Department of Physical Chemistry and Electrochemistry, University of Milan, Italy}

\address[leiden]{Leiden Observatory, Leiden University, P.O. Box 9513, 2300 RA Leiden, The Netherlands }

\address[aarhus]{Department of Physics and Astronomy and Interdisciplinary Nanoscience Centre (iNANO), University of Aarhus, DK-8000 Aarhus C, Denmark }
\cortext[cor1]{Corresponding author. Email: cuppen@strw.leidenuniv.nl}

\begin{abstract}
The formation of H$_2$  in the interstellar medium proceeds on the surfaces of silicate or carbonaceous particles. To get a deeper insight of its formation on the latter substrate, this letter focuses on H$_2$ desorption from graphite (0001) in Temperature-Programmed-Desorption Monte-Carlo simulations. The results are compared to  experimental  results which show two main peaks and an intermediate shoulder for high initial coverage. The simulation program  includes barriers obtained by \emph{ab-initio} methods and is further optimised to match two independent experimental observations. The simulations reproduce the two experimental observed desorption peaks. Additionally, a possible origin of the intermediate peak is given.
\end{abstract}

\begin{keyword}
%% keywords here, in the form: keyword \sep keyword
Temperature Programmed Desorption \sep H$_2$ formation \sep Interstellar Medium \sep Graphite
%% PACS codes here, in the form: \PACS code \sep code

%% MSC codes here, in the form: \MSC code \sep code
%% or \MSC[:2008] code \sep code (:2000 is the default)

\end{keyword}

\end{frontmatter}
%%\linenumbers

%% main text
\section{Introduction}

H$_2$ is found to be the most abundant gas component in the interstellar medium. Formation via simple radiative association is not efficient enough to explain its abundance and already in 1949 it was suggested that surface processes on dust particles could play a role in its formation~\cite{Hulst:1949}. Since carbonaceous material is one of the main constituents of interstellar grains, hydrogen adsorbed on a graphite surface is  often used as a model system.  

H atoms can chemisorb or physisorb on this surface.  In warm regions, with $T>100$~K, the residence time of physisorbed atoms on the surface is too short compared to the flux for two atoms to meet and therefore chemisorbed atoms are believed to play a significant role. Eley-Rideal mechanisms have been largely theoretically studied~\cite{Meijer:2001, Sha:2002a, Morisset:2004, Martinazzo:2006}. 

To study adsorption of H atoms and desorption of H$_2$ molecules on the graphite surface experimentally, the Temperature Programmed Desorption (TPD) technique is successfully applied~\cite{Zecho:2002a}. This technique involves two phases. First, the surface is exposed for a set amount of time to a flux of particles to reach a certain coverage. Then the surface is heated and the desorbing species are measured mass spectrometrically as a function of temperature. In the case of the H(D)/C system, H$_2$ (D$_2$) desorption has been measured and two peaks have been obtained by Zecho \emph{et al.}~\cite{Zecho:2002a}, at 445(490)~K and 560(580)~K. At higher coverages a shoulder appears on the trailing edge of the first desorption peak at intermediate temperatures of 500(540)~K. 

\begin{figure}
\begin{center}
\includegraphics[width=0.35\textwidth]{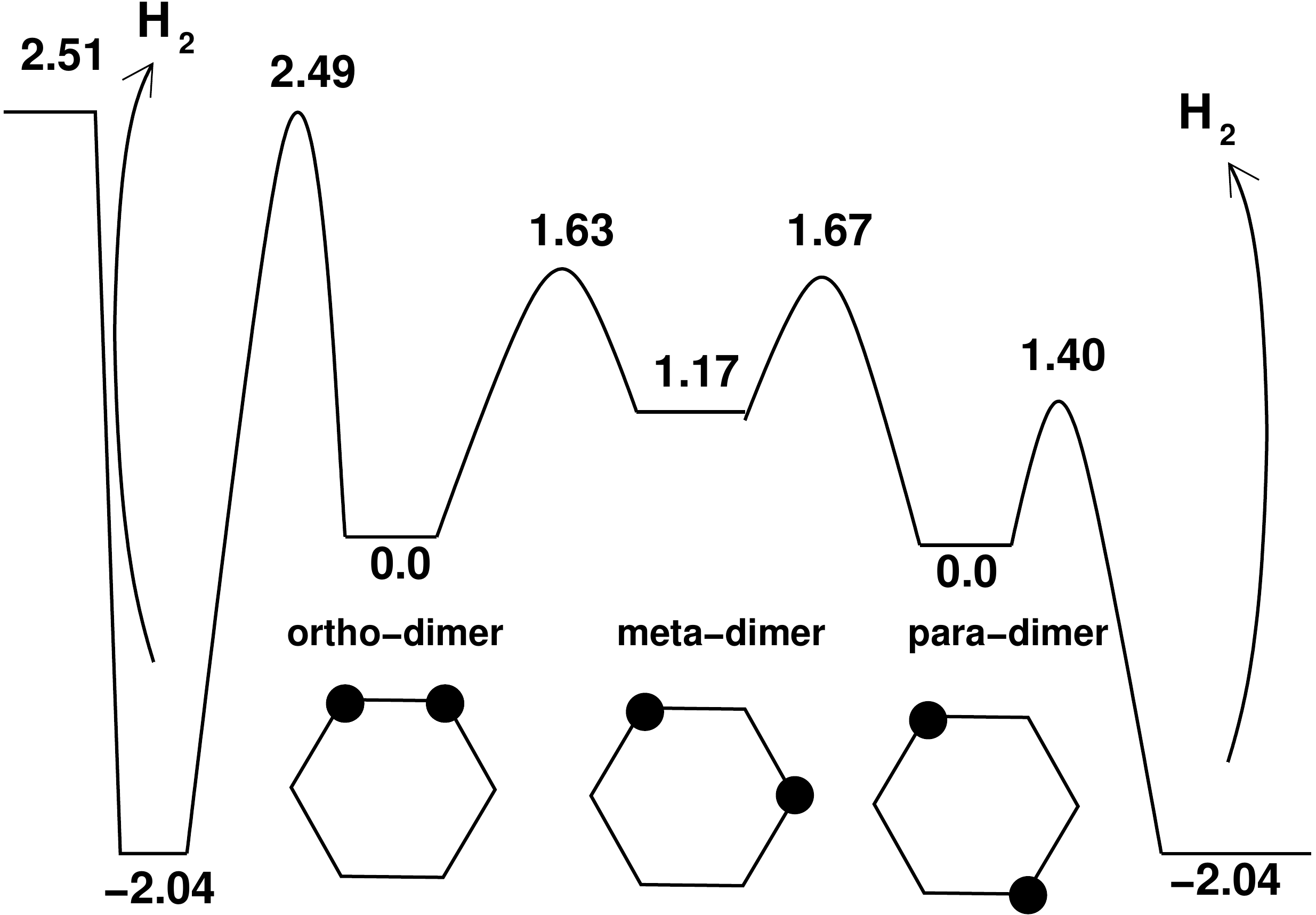}
\end{center}
\caption{Schematic representation of the para-dimer, meta-dimer and ortho-dimer structures 
and the diffusion barriers and H$_2$ formation and desorption barrier from the three dimer configurations (eV), values obtained by DFT calculations~\cite{Hornekaer:2006I}.}
\label{energies}
\end{figure}

The presence of the peak at 560~K has been recently explained by Hornek{\ae}r \emph{et al.}\cite{Hornekaer:2006I}. Their hypothesis involves the presence of dimer structures on the surface. It has been experimentally and theoretically shown that  H atoms can group into stable cluster configurations such as the ortho and para-dimer structures shown in Fig.~\ref{energies}~\cite{Hornekaer:2006I,Ferro:2003,Miura:2003,Rougeau:2006,Andree2006}. The configurations are here schematically shown by the solid circles representing the hydrogen atoms on top of the hexagonal graphite structure. In reality, the hydrogen atoms can be slightly displaced in the xy-plane with respect to the carbon atom it is bound to. The combined binding energy of these clusters is found to be larger than the addition of individual monomer contributions. { Density Functional Theory (DFT) calculations show that the addition of more hydrogen atoms, resulting in trimer and tetramer structures, can further increase the binding energy of the total configuration, where the arrangement of hydrogen atoms determines to which extend this occurs.} Diffusion from an ortho- to a para-dimer structure is possible. Diffusion barriers and the energy required for recombination of H$_2$ molecules are shown in Fig.~\ref{energies} and are taken from Ref.~\cite{Hornekaer:2006I}. H$_2$ formation and desorption from para-dimer configurations occurs at 460~K and accounts for the major contribution to the first peak. As confirmed by scanning tunneling microscopy, ortho-dimers remain longer on the surface because the energy needed for H$_2$ formation and desorption from that configuration is 2.49~eV (1.09~eV higher than the para-dimer H$_2$ formation). Since the diffusion barrier to the meta-dimer is only  1.63~eV, ortho-dimers will transform to para-dimers via the  meta structure at 580~K, see Fig.~\ref{energies}. 

The present letter discusses results obtained from a Monte Carlo simulation of the TPD experiment as performed by Zecho \emph{et al.}~\cite{Zecho:2002a}. Hydrogen structures up to tetramers are explicitly included. The contributions to the two main TPD peaks by abstraction from the different H-atom structures and the possible origin of the third peak will be discussed.

\section{The Monte Carlo method}
For a detailed description of the Monte Carlo algorithm { and a discussion of all the input parameters and energetics} we refer to Ref.~\cite{Cuppen:2008}.  The program uses a lattice-gas representation of the system in which the atoms are limited to certain fixed positions following the lattice. The simulation is divided in two parts: the deposition phase and the heating phase. The deposition time and the heating ramp are tunable.  

The first event in the simulation has to be a deposition, the time at which this deposition occurs is determined by
\begin{equation}
t_{\rm dep}=\frac{\ln(X)\sigma}{f}+t
\end{equation}
where $X$ is a random number generated homogeneously between 0 and 1, $t$ is the current time, $\sigma$ is the density of absorption sites and $f$ is the hydrogen atom flux. The deposited atom can either chemisorb or physisorb. Then atoms on the surface can desorb, diffuse or recombine. Each of these possible events has a rate calculated using 
\begin{equation}
R=\nu \exp\left(-\frac{E}{kT}\right)
\end{equation}
where $T$ is the temperature of the surface, $E$ is the activation energy of the process and  $\nu$ is the attempt frequency ($10^{12}$~Hz for physisorbed atoms and $10^{13}$~Hz for chemisorbed atoms). { Tunnelling is not included in the simulations. At the simulated temperatures most processes are either thermally accessible, or if they do have a large barrier, usually the movement of both a hydrogen and a carbon atom is involved, which results in less efficient tunnelling. For extension to lower temperatures, tunnelling should be considered. }

For each H atom the time at which an event occurs is calculated with 
\begin{equation}
t=\frac{\ln(X)}{R_{\rm dif}+R_{\rm des}+R_{\rm rec}}
\end{equation}
where $R_{\rm dif}$, $R_{\rm des}$ and $R_{\rm rec}$ are the diffusion, desorption  and  recombination rates, respectively. During the heating phase the rates become time and temperature dependent. This can be accounted for using Ref.~\cite{Jansen:1995}, if a known linear temperature ramp is used. Here we use a linear ramp of 1~K/s in accordance with the experiment by Zecho \emph{et al.}~\cite{Zecho:2002a}. Clustering of H atoms is considered, { the total binding energy of a cluster depends on the number of hydrogen atoms it contains and their arrangement. The binding energies are obtained from DFT calculations, see Ref.~\cite{Cuppen:2008} for an overview of the used values.} The desorption energy of { an individual} atom depends on the position of its neighbours and is calculated using 
\begin{equation}
E_{\rm des}=-E_{\rm bind}+E_{\rm bind}^{\rm old}+E_{\rm stick}
\end{equation}
where $E_{\rm bind}$ is the binding energy of remaining cluster and $E_{\rm bind}^{\rm old}$ is the binding energy of the cluster before desorption, the difference between these energy values gives the binding energy of the leaving atom, $E_{\rm stick}$ is the sticking energy corresponding to the barrier that the atom should overcome to chemisorb in its position, this energy gives an extra contribution to the total desorption energy. 

{ An analysis of DFT calculations of the barrier for diffusion from one chemisorbed site to the next, $E_{\rm diff}$, and of the difference in binding energy, $\Delta E$, between initial and final configuration show the following correlation
\begin{equation}
E_{\rm diff} = 0.5 \Delta E + 1.04 \textrm{ eV}.
\label{E_diff}
\end{equation}
The correlation is used to describe the diffusion between chemisorption sites. For diffusion between physisorption sites a rate of $10^{13}$~s$^{-1}$ is used, in agreement with the diffusion coefficient obtained in Ref.~\cite{Bonfanti:2007}.

}

A few improvements are made to our simulation program with respect to ref.~\cite{Cuppen:2008}. First, we include a directional bias for the diffusion of physisorbed atoms, since due to their fast mobility the history of their trajectories is not completely lost  before the next diffusion step.
% and their behaviour is not fully Markovian. 
Originally a physisorbed atom could move in six directions on the hexagonal lattice with equal probability. Now the six directions have different weights with the highest probability for the atom to continue in its original direction. Deviations of more than 60$^{\circ}$ are excluded.  Although this directional bias is probably a more accurate description of the diffusion process, it has very little effect on the final results.  Only if one direction is allowed a lower H$_2$ formation is obtained. 

The trimer sticking barrier has been changed. According to theoretical calculations~\cite{Hornekaer:2006II,Casolo:2009}, the barrier for the chemisorption of an atom close to a dimer is set equal to the barrier for the monomer chemisorption. This choice leads to a larger amount of dimers on the surface at the expense of trimer structures.
 
Finally, two more trimer configurations have been included to complete the group of structures considered. Both trimers consist of an ortho-dimer plus an additional H atom. Since their binding energies are not known, the monomer plus the ortho-dimer energy is used. The addition of these trimers to the list of considered structures leads to the reduction of these configurations due to the reduced sticking barrier with respect to dimers.

The program has several input parameters that are unknown or uncertain. The standard values used in the present letter are the settings obtained in Ref.~\cite{Cuppen:2008} which best reproduced two other independent experimental observations. The dimer energies are taken from Ref.~\cite{Hornekaer:2006I} as explained in Section~\ref{third}. Tests of the sensitivity of the model results on these parameters, however, show that only the peak heights are affected and not the peak positions. The steering parameter, $s$, the thermalization parameters, $T_{\rm{gas}}$ and $B$, the hopping rate of physisorbed atoms and the  chemisorption binding energies are varied for this purpose.  The binding energies are not changed relative to each other. The effect of relative changes will be discussed in Section \ref{third}. 

\section{TPD simulation results}
\begin{figure}
\begin{center}
\includegraphics[width=0.45\textwidth]{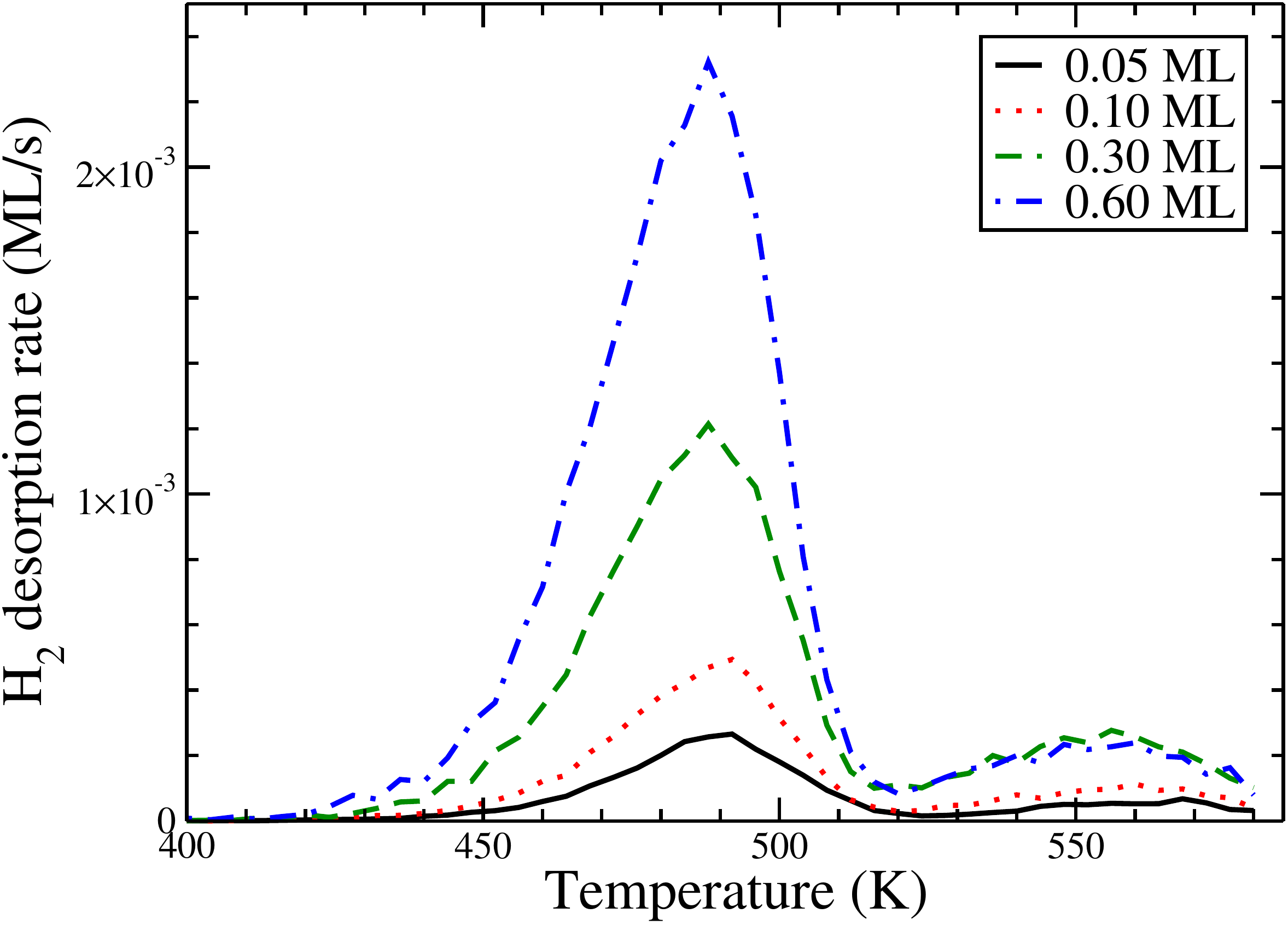}
\end{center}
\caption{Results of the simulation of the TPD experiment, different lines correspond to different initial exposures.}
\label{TPD}
\end{figure}

Figure \ref{TPD} shows the results of the simulation of the TPD experiment for four different initial exposures. It plots the H$_2$ desorption rate in ML as a function of temperature. Each curve is a composition of 15 to 100 independent simulations using a $200 \times 200$ grid for different seeds, depending on the initial exposure{, which are given in ML. Since the sticking probability is lower than one, the resulting coverage is lower.} Two H$_{2}$ desorption peaks are obtained at a temperature of 490~K and 560~K. The first temperature is too high to reproduce the experimental H$_2$ desorption peak of 445~K, but it is in better agreement with the experimental D$_2$ peak. This could be explained by the difference in zero-point energy between H and D which is not taken into account here. The DFT determined binding energies are not explicitly reduced with zero-point energies, which would lead to lower peak temperatures. The effect is expected to be larger for hydrogen than deuterium.

{ The curves show a close to first-order desorption behaviour in agreement with the experiments. One would expect the TPD spectra to show second-order behaviour if the H$_2$ formation reaction is diffusion limited. This is not the case here, the dimers from which the molecules desorb are already formed at low temperature and the desorption of the dimers occurs then first-order.}

Both simulated peaks grow in intensity with increasing initial exposure, again in agreement with experimental data, except for the high temperature desorption peak at 0.6~ML initial exposure. We will return to this in Section \ref{third}. Desorption of H atoms is mainly obtained from mono\-mers, 97.3\% of the total monomer desorption, while  minimal desorption from more complex structures is observed. Desorption from trimers is significantly higher than from clusters with an even number of atoms.

\subsection{Contributions to desorption at 490~K}
\begin{figure}
\begin{center}
\includegraphics[width=0.45\textwidth]{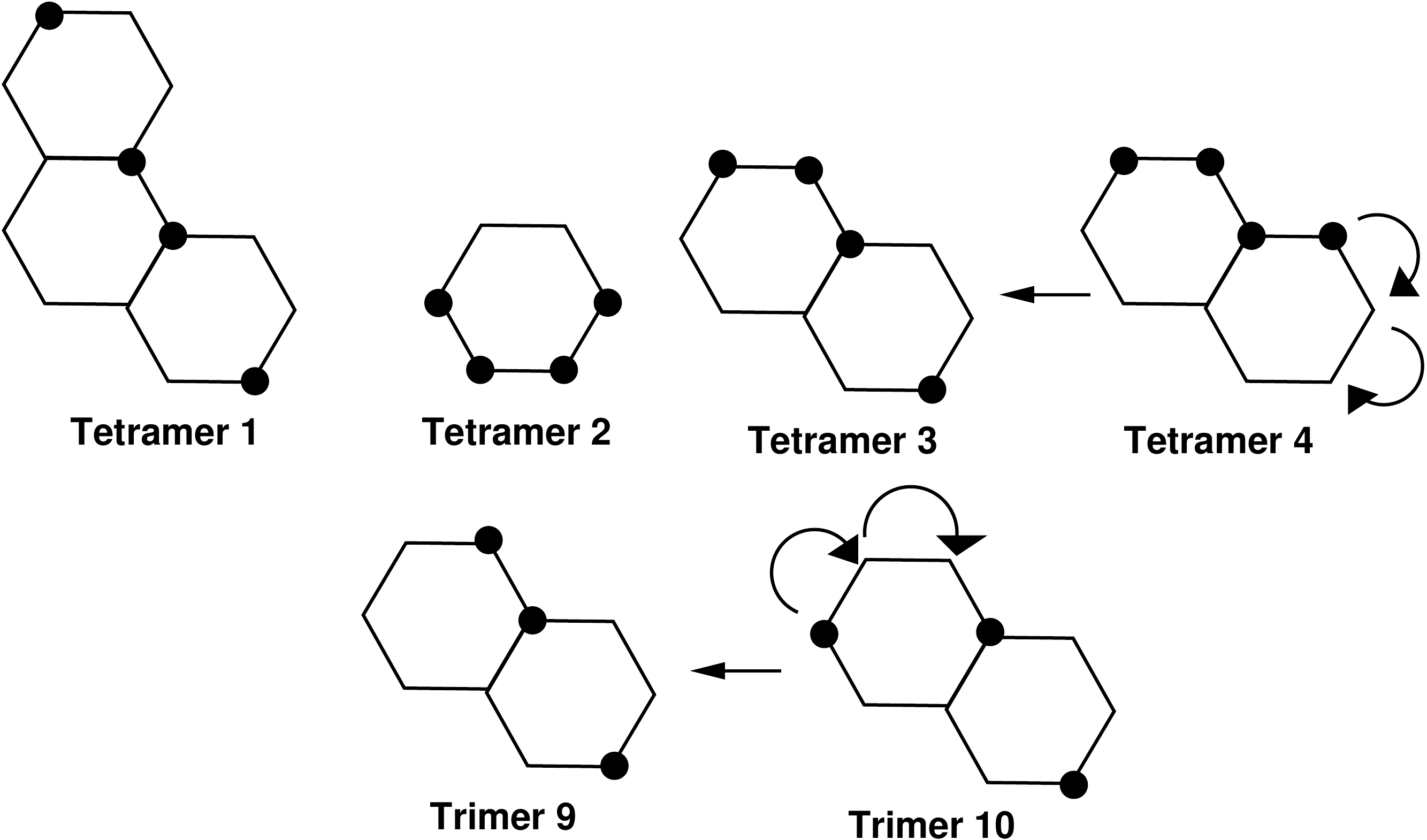}
\end{center}
\caption{Schematic representation of the most frequent H-atom structures. Spheres represent the chemisorbed hydrogen atoms on the graphitic rings.
Some important diffusion processes are indicated as well.}
\label{clusters}
\end{figure}

Clusters of H atoms up to tetramers are taken into account and are labelled by increasing number in order of their stability as shown in Fig.~\ref{clusters}. Thus, tetramer 1 is the most stable structure, whereas trimer 10 has the lowest binding energies of the configurations plotted in this figure. The evolution of the abundance of the individual configurations as a function of temperature and the individual contributions to the TPD spectrum are plotted in Fig.~\ref{config}. Panels (a) and (b) are for a relatively low initial exposure of 0.1 ML and are discussed first. Panel (a) shows that para-dimers and trimer 9 are the dominant structures for these low coverages below a temperature of 490~K. We refer to Fig.~\ref{surface}a for a snapshot of such a surface at 400~K. Trimer 9 contains both para and ortho configuration. Its abundance increases between 360~K and 440~K at the expense of trimer 10 because of diffusion from trimer 10 via trimer 11 as indicated by the arrows in Fig.~\ref{clusters}. Both trimer 9 and the para-dimers disappear at 490~K due to H$_2$ desorption from the two hydrogens at para-postions. Panel (b) clearly shows  that the first peak is indeed due to H$_{2}$ formation and desorption from many configurations including the para-dimer and trimer 9. Two other contributors, tetramers 2 and 3, contain both a para and an ortho-dimer. When H$_2$ desorbs from the two H atoms in para-dimer position two H atoms in the ortho-dimer position are left behind on the surface. This explains the increase in the ortho-dimer abundance between 450~K and 500~K. 

Panels (c) and (d) of Fig.~\ref{config} paint roughly the same picture for an exposure of 0.3~ML, with the difference that more complex structures become dominant on the surface. Tetramers 1, 2 and 3 are now the most important structures. All contain para-dimers and therefore contribute to the H$_2$ desorption at 490~K. Remaining monomers upon H$_2$ desorption, in the case of trimer 9 for instance, immediately desorb.

\subsection{Contributions to desorption at 560~K}
\begin{figure}
\begin{center}
\includegraphics[width=0.48\textwidth]{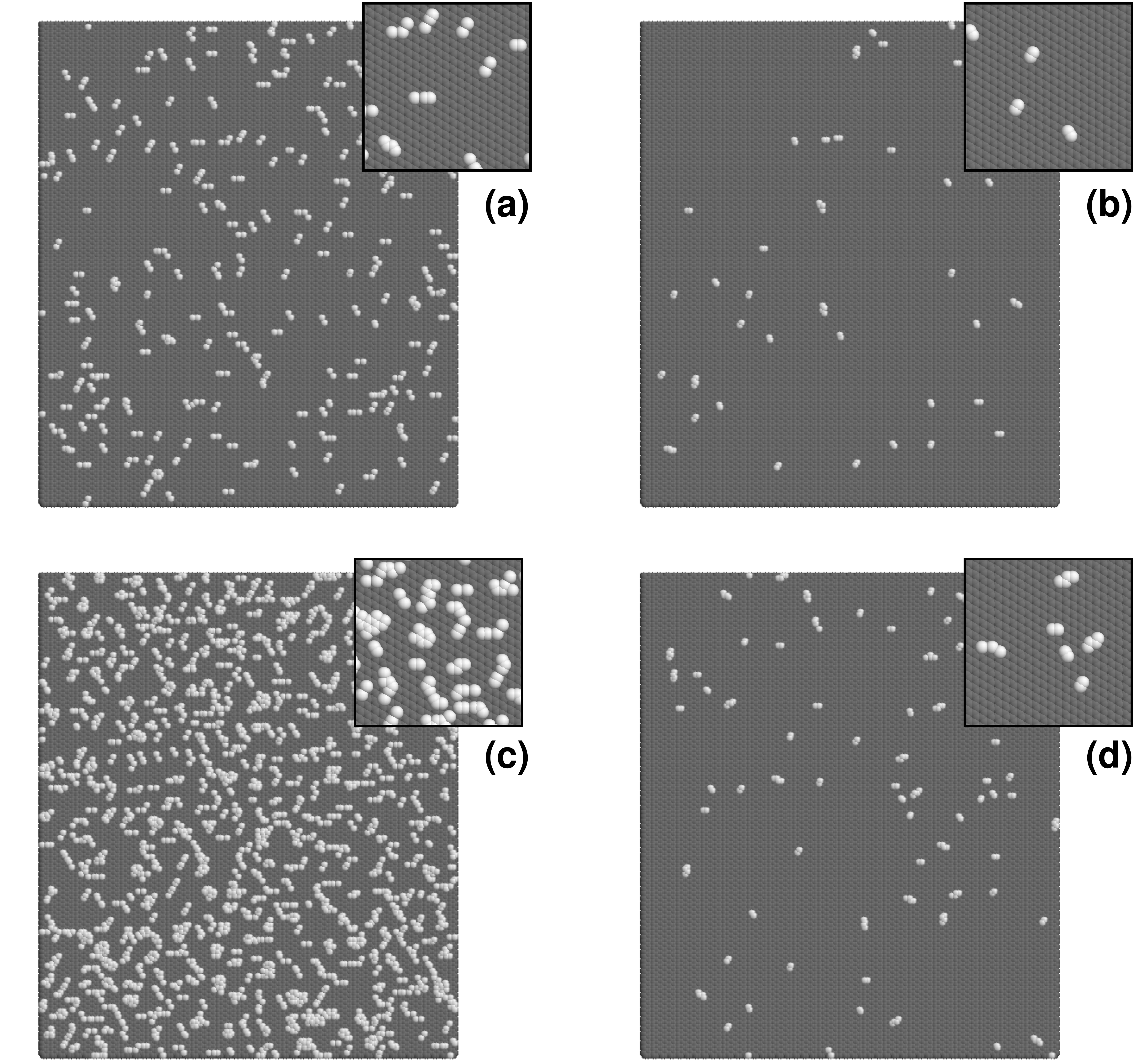}
\end{center}
\caption{H atoms on the graphite surface ($240 \times 210$~\AA) after an exposure of 0.1~ML at (a) 420~K and (b) 530~K and for an exposure of 0.6~ML at (c) 420~K and at (d) 530~K. The insets show small details of the surface.}
\label{surface}
\end{figure}

Figure~\ref{config}b shows that the desorption at 580~K is mainly due to para-dimers although the surface abundance of these structures is negligible at these temperatures. Once the surface is heated to 490~K (Fig.~\ref{surface}b), only ortho-dimers and tetramer 4 configurations remain. Therefore, the diffusion between configurations has been studied to get a better understanding of the exact desorption mechanism. The most important diffusion process is the ortho to para via meta-dimer diffusion. All ortho-dimers diffuse to meta for $T>500$~K, 10\% of the meta-dimers diffuse back to ortho, the rest diffuses to para. At the end of the simulation, all ortho-dimers are converted to para-dimers leaving an almost clean surface behind.  

The H$_2$-formation peak at 580~K does not further increase for high exposures, $\geq1.2$~ML. The main reason for this is that larger pentamer and hexamer structures containing para-dimers are now the dominant configurations, as is clearly visible in Fig.~\ref{surface}c, and formation and desorption of H$_2$ occurs mainly at 490~K. In our simulations, configurations involving  more atoms should be explicitly considered to correctly describe the experiment at these high coverages, while only dimers, trimers and tetramers have been included.

\subsection{Possible origin of intermediate desorption}
\label{third}
\begin{figure}
\begin{center}
\includegraphics[width=0.47\textwidth]{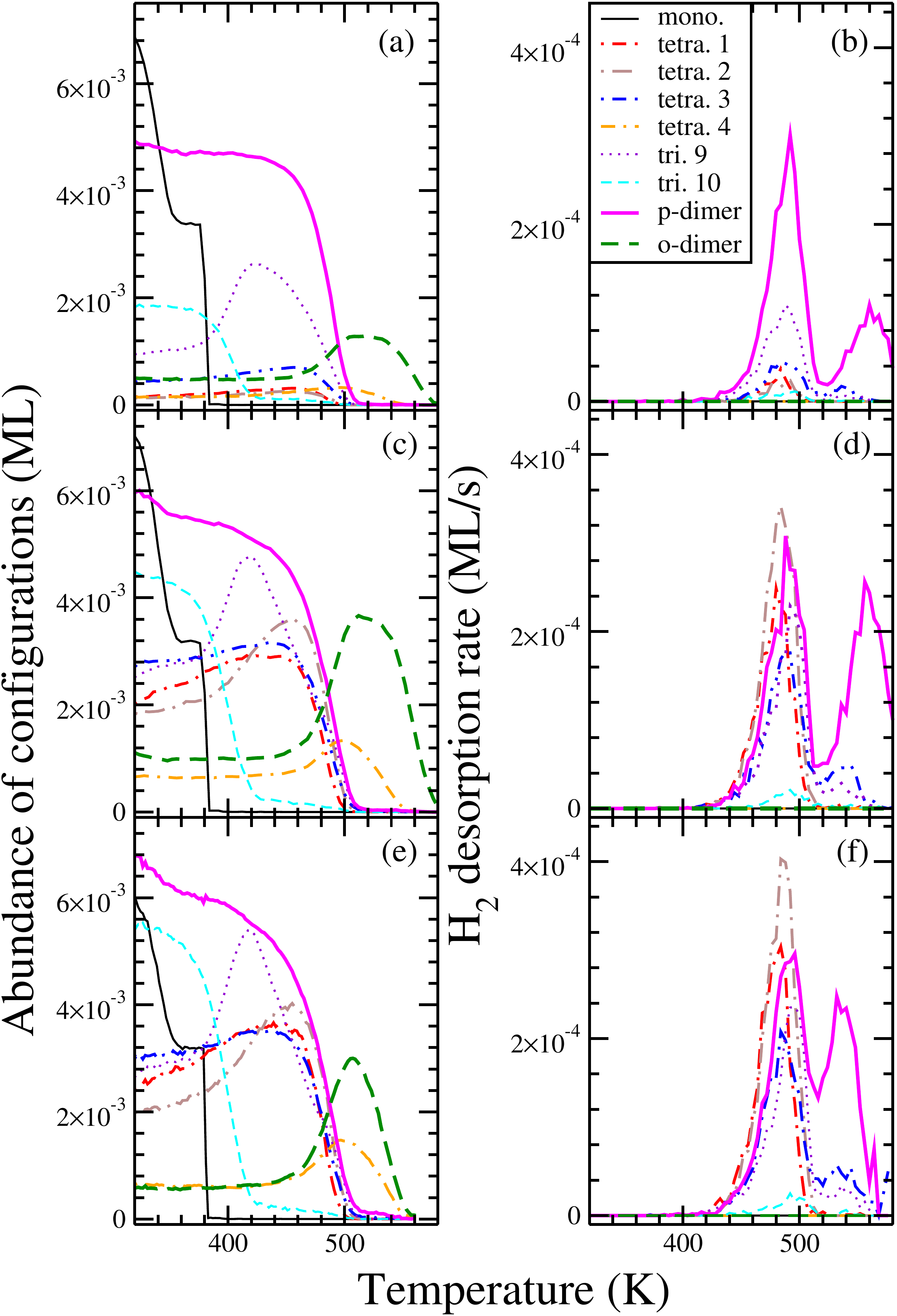}
\end{center}
\caption{(a) Configuration abundances during the heating ramp for an initial exposure of 0.1~ML. (b): TPD simulation 
 results at an exposure of 0.1~ML, detailed contributions of clusters to hydrogen formation are shown.(c) and (d): similar to (a) and (b) for an exposure of 0.3~ML. (d) and (e): similar to (c) and (d) for different dimer energies (see text).} \label{config}\end{figure}

The experimental results~\cite{Zecho:2002a} show a third H$_2$ (D$_2$) abstraction peak at 500 (540)~K. This only occurs at high  initial coverages suggesting more complex structures to be involved. Indeed, Figs.~\ref{config}b and \ref{config}d show small intermediate desorption peaks at 540~K due to tetramer 3 and trimer 9. Tetramer 4 is still abundant after 500~K, then it starts to diffuse via tetramer 8 into tetramer 3 that contains two H atoms in the para-dimer position. Para-dimers  can then recombine giving a contribution to the second peak. The desorption of trimer 9 configurations also originates from tetramer 4, where one of the atoms desorbs leaving a trimer behind that diffuses to the most stable trimer configuration, number 9, before it desorbs as H$_2$.  

The temperature at which this diffusion occurs is determined by the difference in binding energy between the initial configuration and its target. These binding energies are all based on DFT calculations. Different DFT studies result in different relative binding energies, depending on the basis set, functional, and size of the supercell. Here we have used the dimer energies from Ref.~\cite{Hornekaer:2006I} which uses a rather large supercell of 50 atoms. The dimer energies used in our previous paper were based on Ref.~\cite{Hornekaer:2006II}. These calculations use a 32 atom superscell and are performed using PW91, the ones in Ref.~\cite{Hornekaer:2006I} use RPBE. If these dimer energies are implemented, the diffusion barrier for ortho to meta-dimers decreases to 1.54~eV which is exactly the same as the diffusion barrier of tetramer 4. Indeed, the desorption peak at 560~K then overlaps with the H$_2$ desorption at 540~K as is shown in Fig.~\ref{config}e-f. This shows that exact positions of the desorption peaks are very sensitive to the relative binding energies of the configurations and that firm conclusions about which configurations are responsable for the intermediate peak, can only be made if more accurate binding energy calculations are available. Furthermore, scanning tunneling microscopy imaging of the surface at higher coverage at these intermediate temperatures could help confirm/identify the key configurations.

\section{Comparison with previous work}
Dumont \emph{et al.}~\cite{Dumont:2008} performed a similar study in which they simulated the TPD experiment by Zecho \emph{et al.}~\cite{Zecho:2002a} using a Monte Carlo algorithm.  Our explanation for the two TPD peaks agrees with theirs even if we used slightly different energy values, but there are also some differences.  

In their implementation they allowed absorption, desorption and dimerization but they did not include diffusion of physisorbed H atoms on the surface. In a previous paper we found this to be a dominant  mechanism for the formation of clusters~\cite{Cuppen:2008}. 

In the current work the formation of clusters of more than two hydrogen atoms is included explicitly allowing to reproduce  the intermediate peak and to study a larger range of surface coverages. Consequently, the surface snapshots presented by the French group show only dimer configurations whereas the current work shows more complexity in agreement with the experiment~\cite{Hornekaer:2006I}. In their simulations they find an ortho- to para-dimer ratio of 0.5 independent of the experimental conditions. In our simulations an ortho/para ratio of one is obtained at 150~K if a zero sticking barrier for ortho-dimers and for para-dimers is used. In the standard situation ($E_{\rm stick}=0.035$~eV for ortho-dimers~\cite{Casolo:2009}) the ratio is~0.05, this result does not significantly change if another non-zero barrier for chemisorption in the ortho position is set. The ortho/para ratio reaches a value of 0.2 at an H-atom exposure of $\sim 0.01$~ML  for $T_{\rm{grain}}=298$~K and $T_{\rm{gas}}=2200$~K, which compares to the experimental conditions of Ref.~\cite{Hornekaer:2006I} where an experimental ratio of 0.15 was observed. 

Finally Dumont \emph{et al.}~\cite{Dumont:2008} obtained H desorption from ortho-dimer configuration while in our model H desorption occurs  almost exclusively  from monomers and trimers.

\section{Conclusions}
To summarise, Monte Carlo simulations of TPD of H$_2$ from a graphite surface are presented. Two TPD peaks have been obtained for different initial H-atom coverages in agreement with experimental results. Clusters up to four chemisorbed H atoms have been explicitly included. Their abundances on the surface and their contributions to H$_2$ formation have been studied as a function of temperature confirming that the H$_2$ formation and desorption at 490~K is due to para-dimers whereas the peak at 580~K originates from ortho-dimer  diffusion to para-dimers leading to H$_2$ abstraction; moreover similar processes including trimer and tertamer structures have been found that can be the origin of the intermediate desorption peak appearing at higher coverages. 

The main mechanism for H$_2$ formation in a TPD setting appears to be \emph{via} complex structures like dimers, trimer and tetramers, which all contribute to H$_2$ formation at different temperatures. To come back to interstellar conditions, we expect some of the hydrogen molecules to be formed upon H atom sticking where some of the excess energy in the sticking process is used to overcome the barrier for H$_2$ formation and desorption. If the cloud is at too low temperatures for the remaining complex structures to desorb, stochastic heating of the carbonaceous grains by UV photons can facilitate this process. Detailed simulations under these conditions should reveal the dominant process.
%% The Appendices part is started with the command \appendix;
%% appendix sections are then done as normal sections
%% \appendix

%% \section{}
%% \label{}

\section{Acknowledgments}
EG would like to thank the Lifelong Learning Program from the European Union for her stay at the Leiden University. HC is supported by the Netherlands Organization for Scientific Research (NWO) and the Leiden Observatory. LH
acknowledges support from the Danish Natural Science Research Foundation.

\end{document}